\documentclass[aip,jcp,twocolumn,reprint,superscriptaddress]{revtex4-1}

\usepackage[utf8]{inputenc}

\usepackage{graphicx}
\usepackage{braket}
\usepackage{amsbsy}
\usepackage{amsmath}
\usepackage[version=3]{mhchem}
\usepackage{xcolor}
\usepackage{multirow}
\usepackage{xfrac}
\begin{document}

\title{The effect of the Perdew-Zunger self-interaction correction to density functionals on the energetics of small molecules}

\author{Simon Kl\"{u}pfel}
\email[]{simon.kluepfel@gmail.com}
\affiliation{Science Institute of the University of Iceland, Reykjavík, Iceland}
\author{Peter Kl\"{u}pfel}
\affiliation{Science Institute of the University of Iceland, Reykjavík, Iceland}
\affiliation{Laboratoire de Physique Théorique IRSAMC, CNRS, 118 Route de Narbonne, F-31062 Toulouse Cedex 4, France}
\author{Hannes J\'{o}nsson}
\affiliation{Science Institute of the University of Iceland, Reykjavík, Iceland}
\affiliation{Faculty of Physical Sciences, VR-III, University of Iceland, Reykjavík, Iceland}

\begin{abstract}
Self-consistent calculations using the Perdew-Zunger self-interaction correction (PZ-SIC) to local density and gradient dependent energy functionals are presented for the binding energy and equilibrium geometry of small molecules as well as energy barriers of reactions. The effect of the correction is to reduce binding energy and bond lengths and increase activation energy barriers when bond breaking is involved.
The accuracy of the corrected functionals varies strongly, the correction to the binding energy being too weak for the local density approximation but too strong for the gradient dependent functionals considered. 
For the Perdew, Burke, and Ernzerhof (PBE) functional, a scaling of the PZ-SIC by one half gives improved results on average for both binding energy and bond lengths. 
The PZ-SIC does not necessarily give more accurate total energy, 
but it can result in a better cancellation of errors. An essential aspect of these calculations is the use of complex orbitals. 
A restriction to real orbitals leads to less accurate results as was recently shown for atoms [S.~Kl\"upfel, P.~Kl\"upfel, and H.~ J\'onsson, Phys.~Rev.~A {84}, 050501 (2011)]. 
The molecular geometry of radicals can be strongly affected by PZ-SIC.
An incorrect, non-linear structure of the \ce{C2H} radical predicted by PBE is corrected by PZ-SIC.
The  \ce{CH3} radical is correctly predicted to be planar when 
complex orbitals are used, while it is non-planar when the PZ-SIC calculation is restricted to real orbitals. 
\end{abstract}


\maketitle

\newcommand{\rhoup}{\rho^{\uparrow}}
\newcommand{\rhodown}{\rho^{\downarrow}}
\newcommand{\bv}[1]{\boldsymbol{#1}}

\section{Introduction}

Density functional theory (DFT) \cite{Hoh64,Koh65} has become a powerful tool for physicists and chemists to describe the electronic structure of atoms, molecules, and solids. 
While being exact in theory, practical applications of DFT rely on approximations of the exchange-correlation (xc) functional. The local spin density approximation (LSD) \cite{Koh65} is in most cases too crude an approximation for the study of molecular systems. Functionals 
based on the generalized gradient approximation (GGA) \cite{Lan83,Per85a,Per86,Per86_e,Per86a,Per86a_e} allow for a more accurate description of 
the inhomogeneous electron densities
of molecules but also turn out not to be accurate enough in many cases.
The accuracy of GGA functionals for applications of chemical interest can
be further improved by admixture of a fraction of exact exchange 
in the form of hybrid functionals \cite{Bec93a}. The B3LYP \cite{Ste94,Bec93,Bec88,Lee88,Vos80} hybrid functional  
has become widely used in molecular studies.
More recently a number of new functionals have been developed 
to reproduce certain chemical or molecular properties to high accuracy, while being less accurate for others
\cite{Sou07}.

In the spin-unrestricted Kohn-Sham (KS) formalism \cite{Koh65,Bar72} the energy of a system of $N$ electrons in an external potential $v_{\text{ext}}(\bv{r})$ is defined through the spin-densities $\rhoup+\rhodown=\rho$ by
\begin{equation}
E^{\text{KS}}[\rhoup,\rhodown] = T_{\text{s}}[\rhoup,\rhodown] + 
                                   V_{\text{ext}}[\rho] + 
                                   E_{\text{H}}[\rho] + 
                                   E_{\text{xc}}[\rhoup,\rhodown]\ ,
\end{equation}
where $V_{\text{ext}}$ is the electrostatic interaction energy
of the density with the external field and $E_{\text{H}}$ is the Hartree energy, the classical Coulomb repulsion of the charge density with itself. $T_{\text{s}}$ is the kinetic energy of the non-interacting reference system, constructed from the set of KS-orbitals $\lbrace\varphi_{i}\rbrace$ 
to produce the same density as the exact wave function. The exchange-correlation energy, $E_{\text{xc}}$, 
contains all remaining contributions to the exact energy.

The xc-energy can be split into the sum of exchange, $E_{\text{x}}$, and correlation energy, $E_{\text{c}}$. 
For any one-electron density, $\rho^{1}$, the two conditions 
\begin{equation}
E_{\text{x}}[\rho^{1},0] = - E_{\text{H}}[\rho^{1}]
\label{eq:xc_cond_x}
\end{equation}
and
\begin{equation}
E_{\text{c}}[\rho^{1},0] = 0
\label{eq:xc_cond_c}
\end{equation}
are fulfilled by the exact functional. The second condition can be satisfied
by a semi-local form of the correlation functional as the one proposed by 
Lee, Yang, and Parr (LYP) \cite{Lee88}. It is, however, not possible to formulate an exchange functional that can, for any possible $\rho^{1}$, compensate the non-local Hartree energy merely from local information of the density. For approximate semi-local functionals  
condition (\ref{eq:xc_cond_x}), or both conditions, (\ref{eq:xc_cond_x}) and (\ref{eq:xc_cond_c}), are violated. This gives rise to 
a self-interaction error (SIE), 
\begin{equation}
E^{\text{SIE}}[\rho^{1}] = E_{\text{H}}[\rho^{1}] + E_{\text{xc}}[\rho^{1},0]\ .
\label{eq:SIE}
\end{equation}
Perdew and Zunger proposed a self-interaction correction scheme (SIC) in which the SIE of the individual orbitals, as defined by Eq.~(\ref{eq:SIE}), is subtracted from the total energy \cite{Per81}. The Perdew-Zunger self-interaction correction (PZ-SIC) energy functional,
\begin{equation}
E^{\text{PZ-SIC}}[\rho^N] = E^{\text{KS}}[\rhoup,\rhodown] - \sum_{i=1}^{N} E^{\text{SIE}}[\rho_{i}]\ ,
\end{equation}
depends not only on the total spin-densities, but also on the orbital
densities, $\rho^N=(\rho_1,\dots,\rho_N)$ with $\rho_{i}=|\varphi_{i}|^{2}$, and can in principle be applied to
any approximate xc-functional. 

For the exact functional, the correction term vanishes for any many-electron system, 
as can be seen from Eqs.~(\ref{eq:xc_cond_x}) and (\ref{eq:xc_cond_c}).
For approximate functionals, the PZ-SIC is accurate for any one-electron density
but for many-electron systems it is in general only an approximate correction, as the
magnitude of the many-electron self-interaction error does not have to be the sum of the individual SIE terms of Eq.~(\ref{eq:SIE})~\cite{Ruz06}.
While this orbital based estimate cannot be expected to eliminate all
self-interaction for many-electron systems,  it may improve the accuracy of approximate functionals.  
The purpose of the present study is to test the accuracy of PZ-SIC for a few commonly used functionals when applied to molecules. 

Errors in the energy due to approximation of the xc-functional stem from both an incomplete cancellation of the electron self-interaction and an inaccurate description of the inter-electronic interaction. The accuracy of the PZ-SIC does, therefore, depend on the functional approximation it is used with.
From earlier studies, it was concluded that PZ-SIC often overcorrects errors in calculated observables of many-electron systems such as 
equilibrium bond lengths and atomization energy. 
A scaled down modification of the PZ-SIC functional has been proposed in the form of
\begin{equation}
E^{\text{SIC}}[\rhoup,\rhodown] = E^{\text{KS}}[\rhoup,\rhodown] -  \alpha \sum_{i=1}^{N} E^{\text{SIE}}[\rho_{i}]\ ,
\label{eq:scaled_SIC}\end{equation}
and for the Perdew, Burke, and Ernzerhof (PBE) functional a factor
$\alpha$ in the range of 0.4 to 0.5 could improve the description of some observables~\cite{Byl04,Jon11,Val12}.
More elaborate scaling schemes have also been proposed 
\cite{Vyd06,Vyd06a}. 

A recent study of atoms with PZ-SIC showed that the total energy can be lowered significantly by allowing the orbitals to be complex functions~\cite{Klu11}.
This improved significantly the results when 
PZ-SIC was applied to the PBE functional, while a restriction to real orbitals led to much larger errors than those of the uncorrected functional\cite{Vyd04}.
According to the variational principle, the addition of an imaginary component to the orbitals can only lower
the total energy. The effect on equilibrium geometry or energy differences such as atomization energy or energy barriers
is not monotonous in a similar way.
To our knowledge, previously published fully variational studies of the energetics of molecules using
stationary PZ-SIC have exclusively been based on real orbitals. 
Only recently complex orbitals were used in a study focusing on bond-lengths 
of molecules within an approximate Kohn-Sham interpretation of the PZ-SIC \cite{Hof12}.

We present results on the ground state geometry and atomization energy of a set of 17 molecules, the equilibrium structure of two `problematic' radicals and the energy barrier of four reactions. We studied the effect of SIC using three different functionals:
the local spin density approximation (using Slater exchange \cite{Blo29,Dir30,Sla51} and the Perdew-Wang parameterization of correlation \cite{Per92}, SPW92), the two generalized gradient approximations of Perdew, Burke, and Ernzerhof \citep{Per96}, and Becke's exchange functional \cite{Bec88} and correlation of Lee, Yang, and Parr~\cite{Lee88} (BLYP). 
The results of self-consistent calculations using PZ-SIC (SIC) as well as 
the scaled down modification with a factor of one half (SIC/2) are presented. 
For comparison, calculations using the two hybrid functionals, PBE0 \cite{Ada99} and B3LYP \cite{Ste94}, were 
also carried out, as well as less accurate SIC calculations using real orbitals.


\section{Computational method}

The energy minimum with respect to variation of the orbitals under the constraint of orthonormality is described by the two sets of equations \cite{Har83, Ped84}, 
\begin{equation}
\hat{H}_{i} \varphi_{i} = \sum_{i=1}^{N} \lambda_{ji} \varphi_{j}\quad \text{and}\quad 
\lambda_{ji}=\lambda^{*}_{ij}\ ,
\label{eq:SIC_min_cond}
\end{equation}
where the orbital-specific Hamiltonians are defined by
\begin{equation}
\hat{H}_{i} \varphi_{i}(\bv{r})=\frac{\delta E^{\text{SIC}}}{\delta \varphi^{*}_{i}(\bv{r})}\ .
\label{eq:oper_def}
\end{equation}

The Lagrange multipliers can be determined by projection of Eq.~(\ref{eq:SIC_min_cond}) as $\lambda_{ji}=\braket{\varphi_{j}|\hat{H}_{i}|\varphi_{i}}$. In contrast to semi-local functionals, the matrix of Lagrange multipliers is not Hermitian for any unitary transformation among the occupied orbitals. The minimum energy is determined both by the space spanned by the set of orbitals and by the linear combination of the orbitals.

The effective potential will not be the same for all orbitals
and this places SIC outside the domain of Kohn-Sham DFT. 
It can be treated as a true Kohn-Sham functional by means of the optimized effective potential method (OEP) \cite{Kor08a, Mes11}, but in many applications the functional is treated in the generalized Kohn-Sham framework, i.e., the energy is minimized with respect to variation of the orbitals, resulting in different potentials for each one of them.

Analytical gradients of the energy of SIC functionals have
been derived \cite{Goe97} and can be used in a direct minimization of the energy.  The efficiency can be improved
by an additional step in the iterative minimization. Before the orbitals are altered according to the energy gradient, the unitary transformation 
that minimizes the SIC energy terms is found. 
By such a `unitary optimization' the convergence rate can be greatly improved~\cite{Mes09,IncollectionKlu12}. 
An efficient algorithm for the unitary optimization has recently been developed~\cite{Klu12_prep}. 

The calculations were carried out with the Gaussian-type orbital based program \textsc{Quantice}~\cite{wwwQuantice}. 
As analytical gradients of the atomic positions are not available in the program, the molecular structure has been optimized manually. A sequence of internuclear distances and angles was sampled and the minimum energy configuration 
found by cubic interpolation. The equilibrium structure was confirmed by comparison of the interpolated energy with the calculated energy at the interpolated geometry. This manual scheme limits the size of the molecules and the
number of structural degrees of freedom that can be optimized in the current version of this software.
In the geometry optimization, all analogous bonds in a molecule
were constrained to have the same length to reduce the computational effort compared to a completely unconstrained structural relaxation.

For all calculations, atom-centered grids of 75 radial shells \cite{Mur96} of a 302-point Lebedev-Laikov grid \cite{Leb76} were combined to form a multicenter integration grid \cite{Bec88b}. The Cartesian representation of correlation-consistent polarized valence-triple-zeta \cite{Dun89,Woo93} (cc-pVTZ) basis sets were used for most calculations. For the reaction barriers involving only hydrogen, a quadruple-zeta basis set \cite{Woo93} (cc-pVQZ) was used. The convergence criterion of the electronic minimization was $10^{-8}\,\text{Ha}^2$ for the norm of the energy gradient. All calculations were using unrestricted orbitals, starting from a random initialization.


\section{PZ-SIC orbitals}

\begin{figure}[t]
\includegraphics[width=0.85\columnwidth]{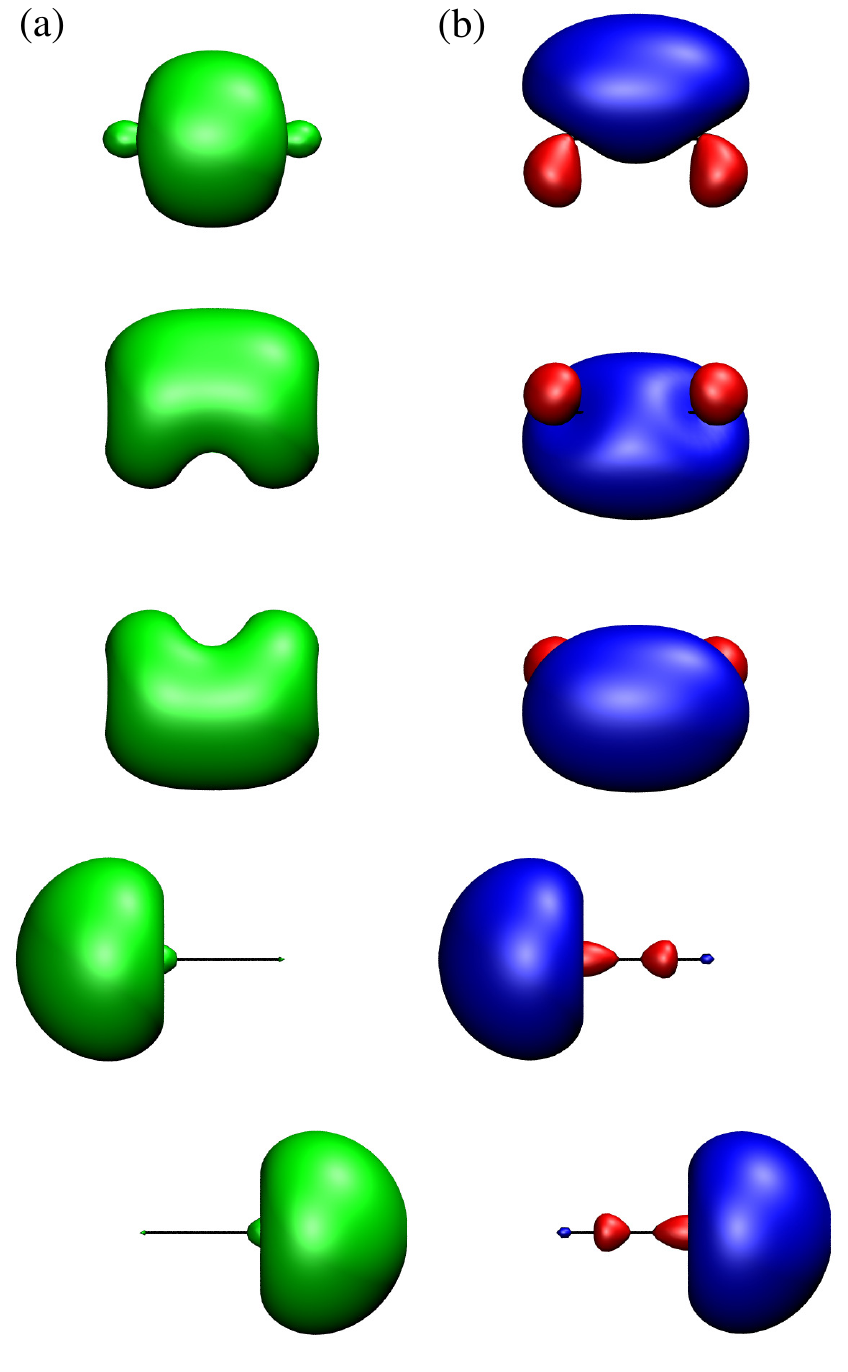}
\caption{Complex (a) and real (b) energy minimizing valence orbitals of \ce{N2} calculated with PBE+SIC.
Only one spin channel is shown, orbitals of the other spin have the same shape.
The top three orbitals represent the triple bond, the bottom two represent lone pairs.
}
\label{fig:N2_orbitals}
\end{figure}

The orbital density dependence of the PZ-SIC energy expression results in a set of well defined `optimal orbitals,'
defined by a unitary transformation of the canonical orbitals. For atomic systems, these orbitals have 
often  been found to resemble $sp^{n}$ hybrid orbitals \cite{Ped88}. 
When the orbitals are allowed to be complex functions, the optimal orbitals 
still resemble hybrid orbitals but can have
significantly different shape and orientation than hybrids of real orbitals. For neon, e.g., real orbitals produce a set of $sp^{3}$ orbitals 
 with tetrahedral orientation, but complex orbitals 
produce a set with tetragonal orientation without nodal surfaces \cite{Klu11}. For molecules, the optimal orbitals often take forms that 
are consistent with `chemical intuition,' i.e. they can be interpreted as lone pairs, single, or
multiple bonds. Also, orbitals consistent with the more `exotic' three-center or banana bonds
can form.  Also here, the shape of the complex orbitals is often rather different from that of the real ones,
but the interpretation above can still be retained in many cases. Figure \ref{fig:N2_orbitals} shows the optimized valence orbitals of one spin channel for \ce{N2}, obtained from PBE+SIC using  complex (a) and real (b) orbitals. 
The real orbitals are of two kinds:
three spatially degenerate orbitals  that add up to a triple bond, and two lone pairs. 
The triple bond is built up from two different kinds of orbitals. 
One orbital has the character of a sigma-bond with rotational symmetry about the molecular axis.
The other two orbitals are degenerate and symmetric about a plane going through the molecular axis.
The complex lone pairs do not differ much qualitatively from the real orbitals.
The real orbitals are rather localized and in staggered orientation, 
but the complex orbitals are more delocalized and to a larger extent share the same space.
%


\section{Atomization energy of molecules}

The ground state energy and equilibrium geometry of the molecules \ce{H2}, \ce{LiH}, \ce{Li2}, \ce{LiF}, HF, \ce{N2}, \ce{O2}, \ce{F2}, \ce{P2}, CO, NO, \ce{CO2}, \ce{CH4}, \ce{NH3}, \ce{H2O}, \ce{C2H2}, and triplet methylene, \ce{CH2},
was calculated.
%
\begin{figure}[htb]
\includegraphics[width=0.85\columnwidth]{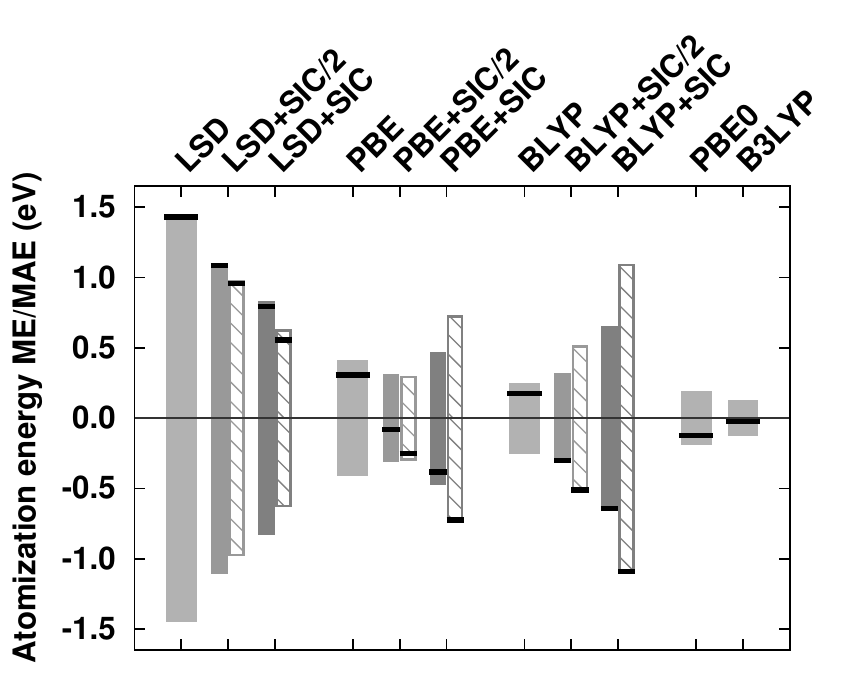}
\caption{Mean error (ME, horizontal black lines) and mean absolute error (MAE, columns) of calculated atomization energy 
compared to experimental values (with zero point energy removed) \cite{Pai05}. 
For comparison, results obtained from calculations restricted to real orbitals are shown by striped columns.
The best overall agreement is obtained with PBE+SIC/2, apart from the hybrid functionals, in particular B3LYP.
}
\label{fig:bond_energy}
\end{figure}

To test the accuracy of the various density functional approximations, 
the predicted atomization energy, i.e., the difference in the total energy of the atoms constituting a molecule and the total energy of the molecule,
was calculated and compared with experimental estimates corrected for zero-point energy \citep{Pai05} or,
in the case of H$_2$, with an accurate theoretical result\citep{Kol65}.
The mean error (ME) and mean absolute error (MAE) for each functional
approximation is shown in Figure \ref{fig:bond_energy}.
For comparison, results of SIC and SIC/2 calculations restricted to real orbitals are also shown. 
The numerical values are listed in Table~\ref{tab:bond_energy},  except for the results obtained with real orbitals.

The LSD energy shows the largest deviation of all functionals, strongly over binding the molecules. 
The errors are reduced by applying SIC (see LSD+SIC), but this only partially eliminates
the errors. The GGA functionals, PBE and BLYP, reduce the errors of LSD significantly, but still predict most molecules 
to be too stable. The binding energy is reduced by SIC for both functionals, but the correction is too large.
For PBE+SIC, the MAE is actually slightly increased and is doubled for BLYP+SIC. The mean deviation is greatly reduced 
by applying SIC scaled by one-half, PBE+SIC/2, while the MAE is just slightly smaller than for the uncorrected functional. BLYP+SIC/2  
gives smaller errors than BLYP+SIC but still predicts the binding energy to be too low.
Calculations using SIC that are restricted to real orbitals predict lower atomization energy on average. 

Vydrov \textit{et al.} studied the effect of PZ-SIC on the heat of formation using several functionals using calculations restricted to
real orbitals \cite{Vyd04}. They concluded that PZ-SIC only improves the results for LSD, while larger deviations are found when 
the correction is used with GGA functionals. 
As shown here, it is important to allow the orbitals to be complex functions.  This reduces the over correction, but does not eliminate it.
Better agreement with the reference data is obtained by scaling the SIC.  No fitting of the scaling factor was carried out, but a factor of one-half chosen
to illustrate the trend. The scaled SIC 
used with the PBE functional gives a smaller mean error than the PBE0 hybrid functional, however, the MAE indicates that some molecules
are over bound while others are too unstable. A systematic under binding is found for  
molecules containing hydrogen, except for \ce{CH2} and \ce{C2H2}. 

\newcommand*{\TPB}[1]{\parbox[c]{1.2cm}{\centering #1}}%
\begin{table*}[t]
\caption{Deviation (in eV) of calculated atomization energy $E_{\text{b}}$ from experiment (with zero point energy removed \cite{Pai05}). For \ce{H2} an accurate result was used as reference \cite{Kol65}. The energy has been calculated for the respective equilibrium geometry.}
\begin{center}
\begin{tabular}{lrrrrrrrrrrrr}
\hline\hline\noalign{\smallskip}
$\Delta E_{\text{b}}$ (eV)  & \TPB{Exp.}  &
\TPB{LSD} & \TPB{+SIC/2} & \TPB{+SIC} &
\TPB{PBE} & \TPB{+SIC/2} & \TPB{+SIC} & \TPB{PBE0} &
\TPB{BLYP} & \TPB{+SIC/2} & \TPB{+SIC} & \TPB{B3LYP} \\ 
\noalign{\smallskip}\hline
\ce{H2  } &  4.75 & 0.16 & 0.18 & 0.22 &-0.21 &-0.26 &-0.30 &-0.22 & 0.00 &-0.04 &-0.07 & 0.03 \\ 
\ce{Li2 } &  1.13 &-0.10 &-0.10 &-0.09 &-0.27 &-0.25 &-0.24 &-0.30 &-0.24 &-0.25 &-0.25 &-0.23 \\ 
\ce{LiH } &  2.52 & 0.11 & 0.13 & 0.17 &-0.21 &-0.24 &-0.26 &-0.24 &-0.01 &-0.07 &-0.12 & 0.01 \\ 
\ce{LiF } &  6.03 & 0.73 & 0.41 & 0.14 &-0.01 &-0.43 &-0.78 &-0.34 & 0.06 &-0.41 &-0.80 &-0.11 \\ 
\ce{HF  } &  6.16 & 0.79 & 0.50 & 0.25 &-0.08 &-0.38 &-0.63 &-0.29 &-0.13 &-0.45 &-0.72 &-0.20 \\ 
\ce{N2  } &  9.84 & 1.75 & 1.38 & 1.05 & 0.73 & 0.33 & 0.00 &-0.07 & 0.59 & 0.02 &-0.46 & 0.11 \\ 
\ce{O2  } &  5.12 & 2.49 & 1.51 & 0.65 & 1.15 & 0.27 &-0.45 & 0.30 & 0.81 &-0.22 &-1.06 & 0.27 \\ 
\ce{F2  } &  1.65 & 1.78 & 0.70 &-0.15 & 0.71 &-0.60 &-1.52 &-0.08 & 0.55 &-1.02 &-1.65 & 0.01 \\ 
\ce{P2  } &  5.03 & 1.14 & 1.00 & 0.89 & 0.18 & 0.16 & 0.18 &-0.27 & 0.19 & 0.07 & 0.01 &-0.08 \\ 
\ce{CO  } & 11.32 & 1.66 & 1.16 & 0.70 & 0.37 &-0.04 &-0.39 &-0.23 & 0.07 &-0.47 &-0.92 &-0.23 \\ 
\ce{NO  } &  6.63 & 2.00 & 1.50 & 1.06 & 0.85 & 0.34 &-0.02 & 0.02 & 0.61 &-0.04 &-0.50 & 0.10 \\ 
\ce{CO2 } & 17.00 & 3.55 & 2.36 & 1.29 & 1.12 & 0.21 &-0.54 & 0.01 & 0.42 &-0.70 &-1.59 &-0.15 \\ 
\ce{CH4 } & 18.21 & 1.84 & 1.91 & 2.04 & 0.03 &-0.23 &-0.43 &-0.09 &-0.10 &-0.46 &-0.74 & 0.06 \\ 
\ce{NH3 } & 12.88 & 1.67 & 1.47 & 1.33 & 0.16 &-0.34 &-0.74 &-0.14 & 0.14 &-0.41 &-0.84 & 0.11 \\ 
\ce{H2O } & 10.10 & 1.35 & 0.94 & 0.58 &-0.03 &-0.51 &-0.92 &-0.34 &-0.11 &-0.59 &-0.99 &-0.18 \\ 
\ce{CH2 } &  8.20 & 1.01 & 1.06 & 1.12 & 0.24 & 0.18 & 0.16 & 0.20 & 0.04 &-0.03 &-0.08 & 0.13 \\ 
\ce{C2H2} & 17.52 & 2.41 & 2.34 & 2.29 & 0.49 & 0.40 & 0.35 & 0.02 & 0.07 &-0.06 &-0.13 &-0.03 \\
\noalign{\smallskip\smallskip}
ME        & \dots & 1.43 & 1.09 & 0.80  & 0.31 & -0.08 & -0.38 & -0.12  & 0.17 & -0.30 & -0.64 & -0.02 \\
MAE       & \dots & 1.44 & 1.10 & 0.82  & 0.40 &  0.30 &  0.46 &  0.19  & 0.24 &  0.31 &  0.64 &  0.12 \\
\noalign{\smallskip}\hline\hline
\end{tabular}
\end{center}
\label{tab:bond_energy}
\end{table*}

An extreme case of the under binding obtained from the BLYP+SIC functional is \ce{F2}, which is predicted to be unstable. Also, PBE+SIC 
gives severe underestimation of the binding energy and elongation of the bond
as shown in Table \ref{tab:F2_bond}. The bond energy is overestimated by all uncorrected functionals.
For LSD, SIC/2 and SIC reduce the bond energy, the latter giving a value closer to experiment. For PBE, the correction
greatly reduces the binding energy, resulting in an underestimation already for SIC/2 and predicting a very weakly bound molecule for 
SIC. This effect is even more pronounced for BLYP, where SIC predicts the molecule not to be bound at all. 
Usually, SIC/2 gives results that are intermediate of those obtained by SIC and by the uncorrected functional.
This is not the case, however, for \ce{F2} in the GGA functionals.
The bond length is reduced by SIC/2, but increased when full SIC is applied. 

\begin{figure}[tb]
\includegraphics[width=0.85\columnwidth]{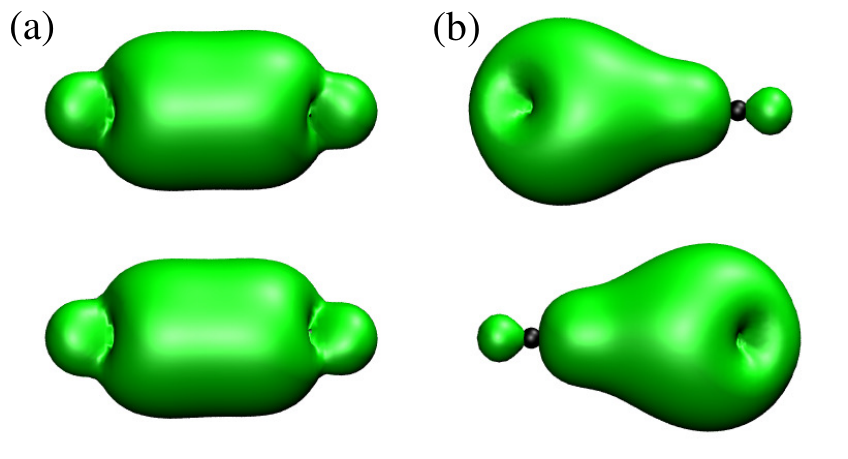}
\caption{Optimized orbitals corresponding to the single bond in \ce{F2} calculated 
with (a) PBE+SIC/2 and (b) PBE+SIC.
Both spin-up and spin-down orbitals are shown. For PBE+SIC/2, the total density is not spin polarized.  
For PBE+SIC, the orbitals are localized to some extent on one of the atoms and the electron density is spin-polarized.}
\label{fig:F2_bond}
\end{figure}

The optimal orbitals obtained using
PBE+SIC/2 and PBE+SIC are qualitatively different, as shown in Figure \ref{fig:F2_bond}. The two
orbitals (one each for spin-up and spin-down) corresponding to the single bond are similar for PBE+SIC/2. 
The slight difference in shape does not result in  
spin polarization of the total density, as it is compensated by the density of the
lone pairs. For PBE+SIC, the orbitals are distorted and localize one on each of the two
nuclei. The total density is spin polarized and the 
electronic structure can be interpreted as an intermediate state towards two separated fluorine atoms. The effect of SIC on the molecule and the single atoms is not balanced. The total energy of the molecule is predicted to be too high
relative to that of the atoms, resulting in a weak bond. For BLYP+SIC the `correction' is unbalanced to such an extent
that the energy of \ce{F2} is above that of the atoms for all nuclear separations. 

Building the bond in a diatomic molecule can be described as a delocalization of atomic hybrid orbitals over both nuclei, accompanied by changes in the shape of the orbitals to maintain orthogonality as well as relaxation
of the orbitals not participating in the bond. The SIC energy of an orbital varies strongly 
depending on how localized it is.
The valence atomic orbitals become more compact 
with increasing atomic number within a row of the periodic table. 
At the same time, the magnitude
by which SIC reduces the atomization energy, increases for all functionals going from \ce{N2} to \ce{O2} to \ce{F2}. Moving from the second to the third row of the periodic table, the valence orbitals become more delocalized, and the changes in atomization energy due to SIC are smaller 
as can be seen by comparing \ce{P2} 
with \ce{N2}. Preliminary results of the binding energy of larger molecules at unrelaxed geometry reveal similar trends.
These trends indicate that the effect of SIC on atomization energy is more pronounced if localized atomic orbitals participate in the bonding. However, without taking into account the changes in the orbital shape and the rearrangement of non-bonding orbitals, such 
a simplified interpretation is insufficient to explain all
the observed effects and a more detailed study is required.

\renewcommand*{\TPB}[1]{\parbox[c]{0.8cm}{\centering #1}}%
\begin{table}[b]
\caption{Atomization energy and equilibrium bond length of \ce{F2}. In BLYP+SIC, the molecule is not stable. The binding energy decreases from the uncorrected functionals to SIC/2 to SIC. The equilibrium bond length, however, changes non-monotonously with the fraction of SIC for the GGA functionals.}
\begin{center}
\begin{tabular}{lrrrclrrr}
\hline\hline\noalign{\smallskip}
\ce{E_b} (eV) & 
\TPB{LSD} & 
\TPB{PBE} & 
\TPB{BLYP} &
\hspace{0.25cm} &
\ce{d_b} (\AA) & 
\TPB{LSD} & 
\TPB{PBE} & 
\TPB{BLYP} \\
\noalign{\smallskip}\hline
Ref.        & 1.65 & 1.65 & 1.65 & & Ref.      & 1.41 & 1.41 & 1.41 \\
Uncorr.  & 3.43 & 2.36 & 2.20 & & Uncorr. & 1.38 & 1.41 & 1.43 \\
SIC/2     & 2.35 & 1.05 & 0.63 & & SIC/2    & 1.33 & 1.36 & 1.38 \\ 
SIC        & 1.50 & 0.13 & 0.00 & & SIC       & 1.30 & 1.44 & \dots  \\
Hybrid   &         & 1.57 & 1.66 & & Hybrid &      & 1.38 & 1.40 \\
\noalign{\smallskip}\hline\hline
\end{tabular}
\end{center}
\label{tab:F2_bond}
\end{table}
%

\begin{figure*}[htbp]
\includegraphics[width=0.85\textwidth]{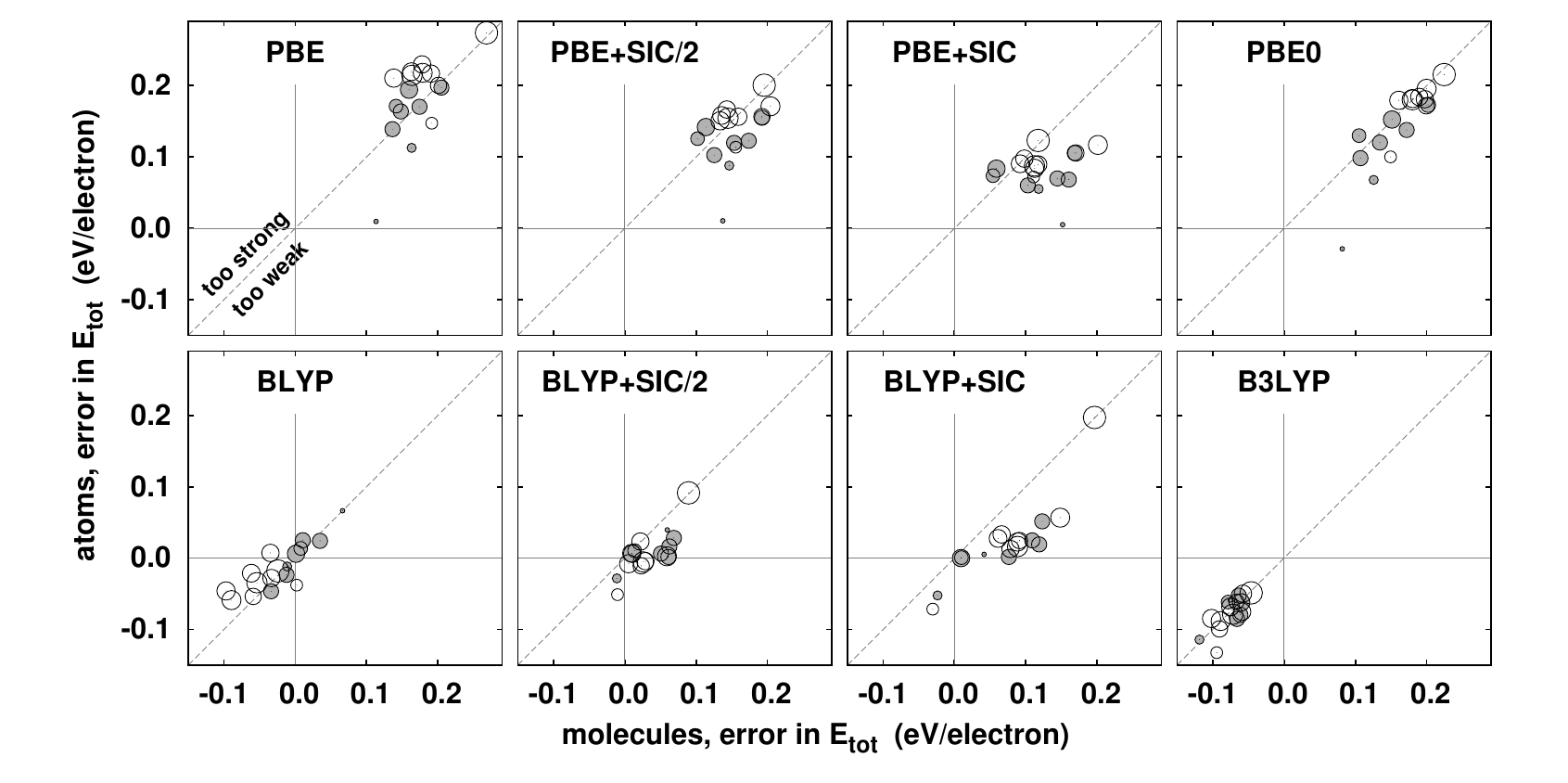}
\caption{Errors per electron in the total energy of the molecules (x-axis) and of the constituent
atoms (y-axis). The diagonal line indicates a cancellation of errors in atomization energy. Systems in the upper left area are over bound, in the lower right area binding energy is too small. Grey points indicate the hydrogen containing molecules. 
}
\label{fig:total_energy}
\end{figure*}

Calculations of observables from energy differences usually are more accurate than the individual total energy values because of partial cancellation of errors.
One source of errors is the limited basis set used in the calculations. This error can in theory be eliminated
by systematically increasing the size of the basis set until a complete basis is reached, or in a more practical way,
until the calculated energy differences do not change significantly. Even if a complete basis set is
used, an error remains from the approximate energy
functional. A functional with large errors in the predicted total energy must be seen as  
too crude an approximation of the exact functional. This, however, does not mean that it can not be a useful
functional for practical calculations. If the functional has a rather constant error per electron for all systems, the
energy difference of systems of the same number of electrons still can be predicted accurately. 

Figure \ref{fig:total_energy} correlates the errors in total energy of the molecules to that of the constituent
atoms. Each circle depicts one molecule, the x-axis shows the error per electron of the molecule, the y-axis that of the atoms. The diagonal dashed line corresponds to a perfect cancellation of errors, for points above or below the line, atomization energy is overestimated or underestimated, respectively. The diameter of the circles indicates the number of electrons in each system. The molecules containing hydrogen are 
indicated by circles filled in grey.

In this representation, the effect of SIC and the admixture of
exact exchange can be studied in more detail than from the atomization energy alone. The total energy of both the atoms and the molecules is overestimated by PBE, around 0.2 eV per electron
 for most molecules. The largest deviation is found for \ce{P2}, the largest system of the test set. With the exception of \ce{H2}
(the smallest point), the errors of the molecules and atoms are similar, but some spread around the diagonal is observed.
Half SIC and full SIC reduce the absolute errors for most systems 
with most points moving closer to the origin. However, the magnitude of the correction is in many cases different for atoms than
molecules. Going from PBE to PBE+SIC/2 to PBE+SIC, 
the points are shifted down and the vertical spread is reduced, which corresponds
to smaller and more similar errors per electron for the atoms. At the same time, however, the horizontal spread increases, indicating a less systematic correction of the errors in the total energy of the
molecules. This can be observed in particular for the molecules containing hydrogen, indicated as grey points. 
When PBE is used, the errors for the molecules are more systematic than the errors for the atoms. 
The opposite trend is observed when PBE+SIC is used.
For PBE0, the absolute magnitude of the errors becomes less systematic for both 
atoms and molecules, the spread is increased along the diagonal. At the same time, the spread perpendicular to the diagonal is reduced, improving the overall cancellation of errors.

For BLYP the total errors are much smaller than for PBE, and the cancellation is slightly better. BLYP+SIC/2 reduces the total errors for many systems, but underestimates the atomization energy. For 
SIC, the points move further below the diagonal but also spread both vertically and horizontally, indicating an unsystematic effect of SIC on the total energy of both atoms and molecules. B3LYP actually increases the magnitude 
of the errors over BLYP, predicting too low energy for all the atoms and molecules. 
The errors are, however, well balanced and cancellation of errors
results in the superior performance of B3LYP with respect to atomization energy, as shown in Figure \ref{fig:bond_energy}. This improved cancellation of errors can to some extent be explained by the origin of this hybrid functional. B3LYP is based on the B3PW91 functional, for
which the parameters had been fit to reference data composed of the total energy of ten atoms and 106 energy differences \cite{Bec93}. 
Such a fitting procedure places less weight on the accuracy of total energy than on   
cancellation of errors. The same parameters are used in the B3LYP functional that, despite its different functional components, predicts energy differences often very accurately while atomic total energy is predicted to be too low.

\section{Equilibrium geometry}
%
\begin{figure}[t]
\includegraphics[width=0.85\columnwidth]{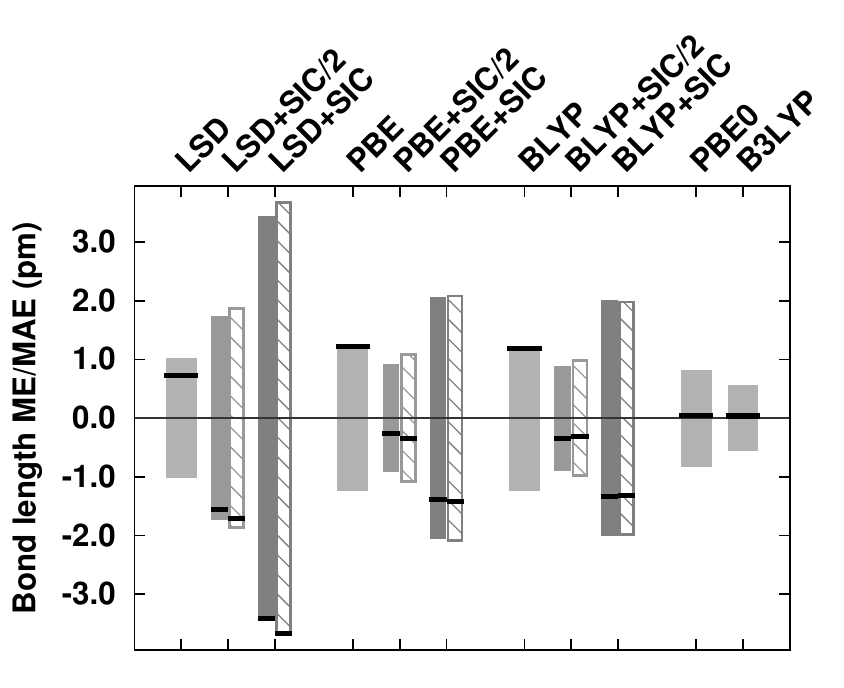}
\caption{Mean error (ME, horizontal lines) and mean absolute error (MAE, bars) of calculated equilibrium bond lengths compared to experimental values 
\cite{wwwNISTCCCBDB}. For SIC/2 and SIC, results obtained using real orbitals are indicated by striped
columns. \ce{F2} was excluded for all functionals, as it is not bound with respect to the atoms for BLYP+SIC.}
\label{fig:bond_length}
\end{figure}
\renewcommand*{\TPB}[1]{\parbox[c]{1.2cm}{\centering #1}}%
\begin{table*}[htbp]
\caption{Deviation (in pm) of calculated bond length $d_{\text{b}}$ from experimentally determined geometry \cite{wwwNISTCCCBDB}.}
\begin{center}
\begin{tabular}{lrrrrrrrrrrrr}
\hline\hline\noalign{\smallskip}
$\Delta d_{\text{b}}$ & \TPB{Exp.}  &
\TPB{LSD} & \TPB{+SIC/2} & \TPB{+SIC} &
\TPB{PBE} & \TPB{+SIC/2} & \TPB{+SIC} & \TPB{PBE0} &
\TPB{BLYP} & \TPB{+SIC/2} & \TPB{+SIC} & \TPB{B3LYP} \\ 
\noalign{\smallskip}\hline
\ce{H2  }     &  74  & 2.5 & 0.5 &-1.4  & 1.0 & 0.0 &-0.9 & 0.5  & 0.6 &-0.4 &-1.2 & 0.2 \\ 
\ce{Li2 }     & 267  & 3.6 & 0.9 &-1.1  & 5.6 & 5.3 & 5.6 & 5.6  & 3.9 & 4.2 & 5.6 & 2.8 \\ 
\ce{LiH }     & 160  & 1.1 &-1.7 &-4.2  & 1.2 &-0.1 &-0.9 & 0.4  & 0.4 &-0.4 &-1.1 &-0.4 \\ 
\ce{LiF }     & 156  &-1.0 &-3.3 &-5.2  & 1.2 &-0.5 &-1.8 &-0.1  & 1.4 &-0.2 &-1.0 & 0.1 \\ 
\ce{HF  }     &  92  & 1.4 &-0.8 &-2.7  & 1.3 &-0.6 &-2.0 & 0.1  & 1.6 &-0.5 &-2.0 & 0.5 \\ 
\ce{N2  }     & 110  &-0.2 &-2.1 &-3.6  & 0.5 &-0.9 &-1.9 &-0.9  & 0.5 &-0.9 &-1.9 &-0.6 \\ 
\ce{O2  }     & 121  &-0.3 &-3.8 &-6.4  & 1.2 &-2.4 &-4.9 &-1.4  & 2.3 &-1.8 &-4.3 &-0.3 \\ 
\ce{F2  }     & 141  &-2.8 &-8.1 &-11.5 & 0.0 &-5.7 & 2.7 &-3.7  & 1.9 &-3.7 & \dots &-1.6 \\ 
\ce{P2  }     & 189  & 0.5 &-2.3 &-4.4  & 1.8 &-0.1 &-1.5 &-0.5  & 2.6 & 0.3 &-1.1 & 0.3 \\ 
\ce{CO  }     & 113  & 0.0 &-1.9 &-3.5  & 0.9 &-0.5 &-1.7 &-0.4  & 1.0 &-0.5 &-1.7 &-0.3 \\ 
\ce{NO  }     & 115  &-0.7 &-3.1 &-5.0  & 0.5 &-1.6 &-3.1 &-1.4  & 0.8 &-1.4 &-3.0 &-0.8 \\ 
\ce{CO2 }     & 116  & 0.1 &-2.2 &-4.0  & 1.0 &-0.8 &-2.2 &-0.5  & 1.1 &-0.8 &-2.1 &-0.2 \\ 
\ce{CH4 }     & 109  & 1.0 &-0.7 &-2.3  & 0.9 & 0.2 &-0.5 & 0.2  & 0.7 &-0.1 &-0.2 & 0.2 \\ 
\ce{NH3 }     & 101  & 1.6 &-0.4 &-2.2  & 1.5 & 0.0 &-1.2 & 0.5  & 1.5 &-0.2 &-1.5 & 0.7 \\ 
\ce{H2O }     &  96  & 1.3 &-1.1 &-2.9  & 1.2 &-0.2 &-1.8 & 0.0  & 1.3 &-0.6 &-2.2 & 0.3 \\ 
\ce{CH2 }     & 109  & 0.4 &-1.6 &-3.5  & 0.0 &-1.2 &-2.3 &-0.7  &-0.2 &-1.5 &-2.4 &-0.8 \\ 
\ce{C2H2} (CC)& 120  &-0.1 &-1.8 &-3.1  & 0.5 &-0.3 &-1.1 &-0.6  & 0.3 &-0.6 &-1.4 &-0.8 \\ 
\ce{C2H2} (CH)& 106  & 1.1 &-0.8 &-2.6  & 0.7 &-0.5 &-1.2 & 0.1  & 0.4 &-0.5 &-1.3 &-0.1 \\ 
\noalign{\smallskip\smallskip}
ME            & \dots   & 0.7 &-1.6 &-3.4  & 1.2 &-0.3 &-1.4 & 0.0  & 1.2 &-0.3 &-1.3 & 0.0 \\ 
MAE           & \dots   & 1.0 & 1.7 & 3.4  & 1.2 & 0.9 & 2.0 & 0.8  & 1.2 & 0.9 & 2.0 & 0.5 \\ 
\hline\hline
\end{tabular}
\end{center}
\label{tab:bond_length}
\end{table*}

The equilibrium geometry of a molecule is found as the minimum of the potential energy surface, determined from the total energy of the molecule at different geometries. Again, a constant error at all geometries will still allow for the prediction of the correct equilibrium structure, while varying errors can result in quantitatively or even qualitatively wrong structures.

Figure \ref{fig:bond_length} shows the mean and mean absolute error for the equilibrium bond lengths of the 
set of molecules and functionals studied, excluding \ce{F2}, as it was shown to be not stable in BLYP+SIC. The  
numerical values are listed in Table~\ref{tab:bond_length}.  The uncorrected 
functionals overestimate the bond lengths on average slightly by $\sim$1pm, the GGA functionals generally overestimate the bond length, whereas some molecules are predicted to have too short bonds by LSD.  
PZ-SIC results in a strong overcorrection. For LSD+SIC all bonds are predicted to be much too short. 
For PBE+SIC and BLYP+SIC, the overcorrection is 
smaller. Here, all bond lengths except that of \ce{Li2} (and \ce{F2}) are predicted to be too short. 

The scaled SIC overcorrects LSD but gives on average improved results for the GGA functionals. Still, as in the case of atomization energy, the mean absolute error shows
large fluctuations. A restriction to real orbitals has less effect on the bond length than on the atomization energy but gives slightly worse results except for 
BLYP+SIC/2 and BLYP+SIC. The hybrid functionals give the highest accuracy.

\begin{figure}[t]
\includegraphics[width=0.85\columnwidth]{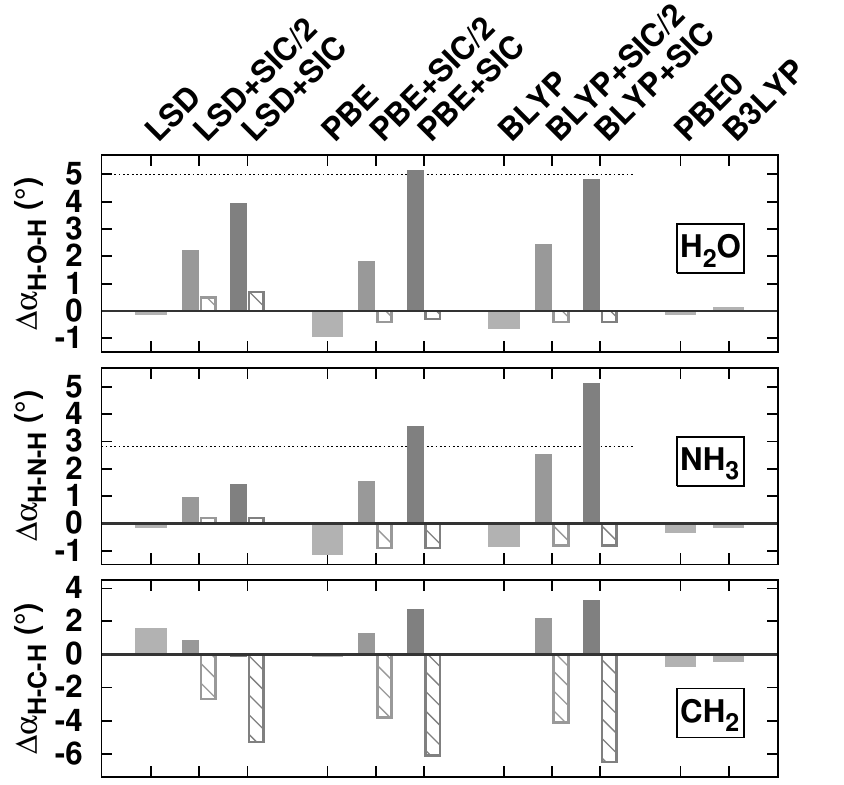}%
\caption{Bond angle deviations for \ce{H2O}, \ce{NH3}, and \ce{CH2}. 
The difference of calculated and experimental \cite{wwwNISTCCCBDB} angles H-X-H are shown for the various functionals.}
\label{fig:bond_angles}
\end{figure}

Figure \ref{fig:bond_angles} shows the deviation of the equilibrium bond angles from experimental geometries for the non-linear molecules. For \ce{H2O} and \ce{NH3}, the angle from LSD and the hybrid functionals are in very good agreement with experiment, while the GGA functionals predict angles 0.5$^\circ$-1.0$^\circ$ too small. SIC/2 and SIC predict larger angles for all functionals, with a monotonous increase from SIC/2 to SIC. For the full correction,
the angle in water is close to that of a regular tetrahedral structure, indicated by a dotted 
line, in ammonia it even exceeds this angle, in particular for BLYP+SIC. 
The localized nature of the optimal orbitals can motivate an interpretation along the lines of valence shell electron pair repulsion (VSEPR) theory. 
The increase in bond angles indicates a relatively stronger interaction between bonding electron
pairs compared to the repulsion between a lone pair and a bond pair. 
In methylene, the bond angle is predicted to be too large by the LSD functional but is quite accurately predicted by 
the GGA functionals. The angle is reduced in LSD+SIC, to good agreement with experiment, while it increases for the GGA+SIC functionals.
 
In all cases, the angles predicted by using real orbitals with SIC are lower than when complex orbitals are used, and 
except for methylene closer to experiment. In contrast to most equilibrium bond lengths, 
restriction to real orbitals has a strong effect on the equilibrium bond angles. 
The hybrid functionals here also give results that are closer to experimental results.

\section{
Structure of molecular radicals}
%
\begin{figure}[t]
\includegraphics[width=0.85\columnwidth]{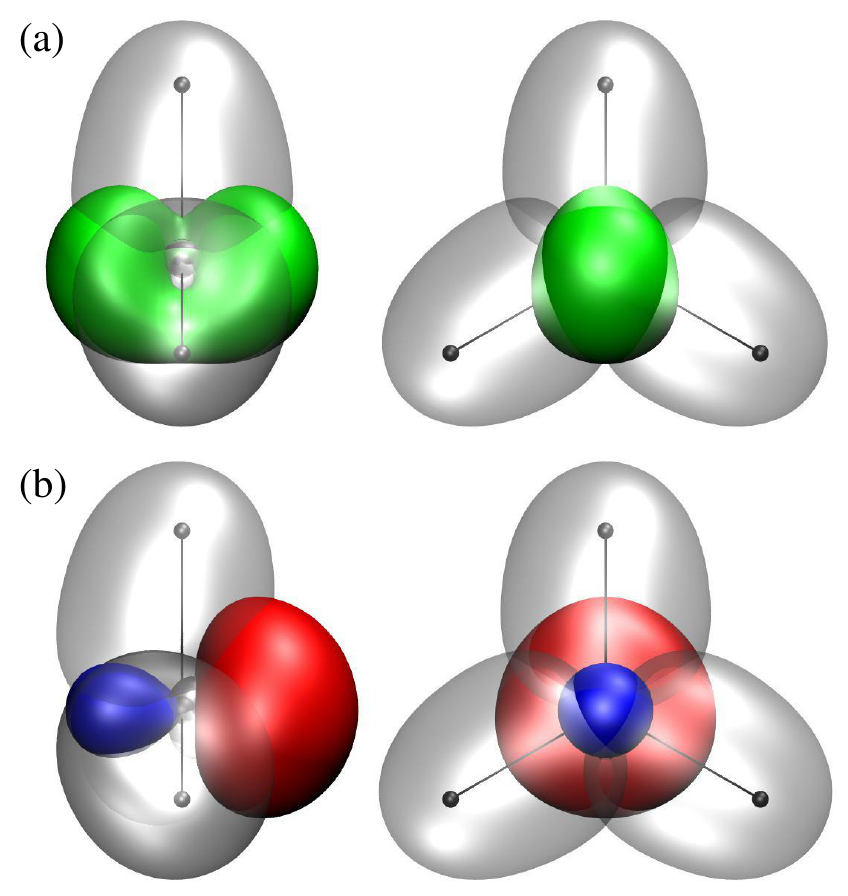}
\caption{(a) Complex  and (b) real PBE+SIC optimized valence orbitals of the planar \ce{CH3} radical. Isosurfaces of the spin-majority valence orbital densities are shown in side 
view (left) and top view (right). The 
orbital of the unpaired electron 
is colored. The complex orbitals have mirror symmetry with respect to the plane. The real orbital for the unpaired electron
has $sp^{3}$ character and the C-H-binding orbitals are out of plane. The arrangement of the real orbitals is not favored and the ground state geometry is predicted to be pyramidal \cite{Gra04a}. Figure by Simon Klüpfel from ‘Implementation and reassessment of the Perdew-Zunger self-interaction correction’, ISBN: 978-9935-9053-8-3. Used under a Creative Commons Attribution license.}
\label{fig:CH3_orbitals}
\end{figure}

Gr\"{a}fenstein \textit{et al.} found
that the equilibrium structure of the \ce{CH3} radical predicted by
BLYP+SIC is non-planar \cite{Gra04a}, in disagreement with both experimental observation \cite{wwwNISTCCCBDB} and ab initio calculations \cite{Gra04a}. We determined the ground state geometry, restricted to $C_{3v}$ symmetry, and found that all the uncorrected functionals predict the proper
planar structure. Results obtained from applying SIC are listed in Table \ref{tab:CH3_angle}
as the energy difference between the equilibrium structure and
the planar structure of lowest energy, and as the out-of-plane angle, i.e., the angle enclosed by a C-H bond and the plane defined by the hydrogen atoms. 
Using real orbitals, the non-planar geometry is preferred in both SIC and SIC/2 for all the functionals.
Both the angle and energy difference increase from SIC/2 to SIC and are larger for the GGA functionals than for LSD. Using complex orbitals, however, the self-interaction corrected GGA functionals predict the geometry to be
qualitatively correct, while it merely reduces the angle and energy difference for LSD, retaining the incorrect pyramidal structure.

\renewcommand*{\TPB}[1]{\parbox[c]{0.8cm}{\centering #1}}%
\begin{table}[b]
\caption{Equilibrium structure of the \ce{CH3} radical. The 'out of plane' angle, $\alpha$, in degrees and energy difference between planar and pyramidal structure, $\Delta E$, in meV is shown for the various SIC functionals for complex (c.) and real (r.) orbitals. The uncorrected functionals all predict the correct planar ground state. 
}
\begin{center}
\begin{tabular}{llcccccccccccccccc}
\hline\hline\noalign{\smallskip}
&    & & \multicolumn{2}{c}{LSD} & \hspace{0.1cm} & \multicolumn{2}{c}{PBE} &  \hspace{0.1cm} & \multicolumn{2}{c}{BLYP}\\
\cline{4-5}\cline{7-8}\cline{10-11}\noalign{\smallskip}
&    & & 
SIC/2 & SIC & & 
SIC/2 & SIC & & 
SIC/2 & SIC  \\
\noalign{\smallskip}\hline\noalign{\smallskip}
$\alpha$ ($^\circ$) 
&\ c. & & 5.0 & 7.2        & & 0.0 & 0.0               & & 0.0 & 0.0  \\
&\ r. & & 6.7 & 8.7        & & 7.1 & 9.1               & & 7.4 & 9.6  \\
\noalign{\smallskip}
$\Delta E$ (meV)
&\ c. & &  8 &  41     & & \dots & \dots                   & & \dots & \dots  \\
&\ r. & & 37 & 109     & & 43 & 119                    & & 51 & 142  \\
\noalign{\smallskip}\hline\hline
\end{tabular}
\end{center}
\label{tab:CH3_angle}
\end{table}

The destabilization of the planar structure by using SIC with real orbitals can be understood from the hybridization of the optimized orbitals. Figure \ref{fig:CH3_orbitals} depicts the complex and real optimized valence orbitals of the spin majority for the planar equilibrium structure predicted by
PBE+SIC. The complex orbitals corresponding to C-H $\sigma$-bonds lie in the plane. The orbital of the unpaired electron
is delocalized symmetrically over both sides of the plane
with an increased density between two of the bonding orbitals. The shape of these two orbitals differs slightly from the shape of the third
one. The total and spin density have, despite the reduced symmetry of the optimal orbitals, the full symmetry of the molecule.

The shape of the real orbitals is quite different. The orbital of the unpaired electron
takes the form of a real $sp^{3}$ hybrid. As the orbitals have to be orthogonal, the binding orbitals are forced into an unnatural shape, `bending' out of plane between carbon and hydrogen. While an $sp^{2}$ configuration would seem more favorable, this would force the unpaired orbital to be an unhybridized 
$p$-orbital, which is higher in energy. The SIC energy contribution of the unpaired orbital is lowered to such an
extent by the hybridization, that it compensates for the higher SIC energy terms
of the `banana-bonds.' The total energy can be lowered further, by moving the hydrogen atoms out of the plane, which results in a geometry in better agreement with the $sp^{3}$ hybridization on the carbon atom. 

For LSD+SIC (not shown), the unpaired optimal orbital appears as an intermediate between the real and complex orbital shown for PBE+SIC. The orbital is partially delocalized but is not symmetric with respect to the plane. For atoms, it was found that extending the variational space to 
complex orbitals lowers the total energy in GGA+SIC more than in LSD+SIC \cite{Klu11}.
This can play a role in the 
equilibrium geometry. The SIC energy of the unpaired and binding orbitals will be affected differently by 
the additional complex degrees of freedom
and a comparison between GGA+SIC and LSD+SIC might give insight into the origin of the insufficient correction found for LSD+SIC.
However, as the bond lengths are different for real and complex orbitals, all terms of the energy functional change, making such an analysis 
more difficult.

\begin{figure}[tb]
\begin{center}
\includegraphics[width=0.85\columnwidth]{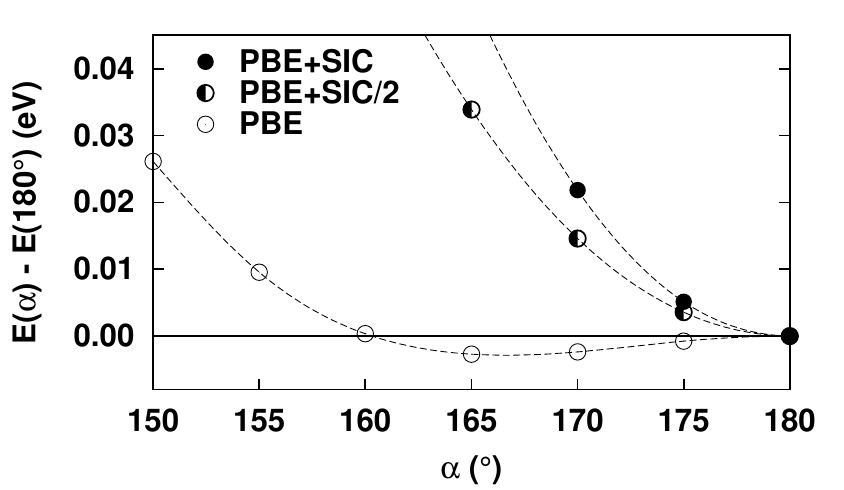}%
\caption{Energy of the bent ethynyl radical relative to the energy of the linear geometry. The bond lengths have been optimized  
for each value of the bond angle. 
The PBE ground state geometry is bent with an angle of $\approx$166$^\circ$ in agreement with previous calculations \cite{Oye12}. PBE+SIC and PBE+SIC/2 both favor the linear geometry in agreement with experiment \cite{wwwNISTCCCBDB}.}
\label{C2H_geometry}
\end{center}
\end{figure}

Recently, Oyeyemi \textit{et al.} found that PBE predicts an incorrect equilibrium structure of the ethynyl radical, \ce{C2H} \cite{Oye12}. 
The structure is not linear, but rather
bent by $\sim$166$^\circ$. This was attributed to an over delocalization of the electron density due to the self-interaction error of approximate functionals. By including exact exchange in the form of hybrid functionals, the correct linear ground state geometry is obtained. 
Figure \ref{C2H_geometry} shows the energy of the bent radical relative to the energy of the linear structure. For all angles, bond lengths have been optimized. For PBE the energy drops slightly when the molecule starts to bend, with an optimal angle in good agreement with the previous study \cite{Oye12}. For PBE+SIC and PBE+SIC/2,
 the linear geometry is favored. The self-interaction correction
corrects the Hartree self-energy, as does exact exchange.
In cases where inaccuracy of approximate functionals stems
from the spurious self-repulsion of the orbitals, (scaled) PZ-SIC  
corrects in a way that is analogous to hybrid functionals 
which include scaled exact exchange.


\section{Reaction barriers}

\begin{table}[b]
\caption{
Energy barrier for four reactions.
For each saddle point, the point group, energy barrier with respect to the reactants, and bond-length are listed. The labels
in bold face are used throughout the text for the saddle point.}
\begin{tabular}{lclcc}
\hline\hline
\multicolumn{1}{l}{Reaction}                    & \hspace{0.2cm} & sym.          & $E^\#$(eV) & $r^\#$(\AA)\\
\hline
\ce{H2 + H  ->}\textbf{\ce{H3}} \ce{-> H  + H2} & & $D_{\infty h}$ & 0.42 & 0.93\\
\ce{H2 + H2 ->}\textbf{\ce{H4}} \ce{-> H2 + H2} & & $D_{4h}$       & 6.42 & 1.23\\
\ce{HF + H  ->}\textbf{\ce{HFH}}\ce{-> H  + HF} & & $D_{\infty h}$ & 2.12 & 1.14\\
\ce{H-NH-H  ->}\textbf{\ce{NH3}}\ce{-> H-HN-H } & & $D_{3h}$       & 0.22 & 0.99\\
\hline\hline
\end{tabular}
\label{tab:barriers}
\end{table}

The energy barrier for four reactions has been calculated to study the effect of SIC on the activation energy.
The energy difference between the reactant minimum and the lowest saddle point of the potential energy surface is calculated. 
The reactions are listed in Table \ref{tab:barriers} as well as the
geometry of the saddle point and the barrier height
evaluated from \textit{ab initio} calculations. For \ce{H4} and \ce{H3},
the saddle point had been calculated using configuration interaction (CI) \cite{Sch99}. The barrier heights
are calculated with respect to the reactants, using the accurate total energy of the hydrogen molecule \cite{Kol65} and the exact energy of the hydrogen atom as reference values. 
For HFH, only the collinear, symmetric saddle point was considered, as a reference for this barrier is available \cite{Ben75}, and the computational effort was too large to find the non-collinear saddle point which is slightly lower in energy \cite{Wad77}. The inversion barrier height of \ce{NH3} 
had been calculated using the coupled cluster method, CCSD(T) \cite{Lin02}. 

\begin{figure}[t]
\includegraphics[width=0.85\columnwidth]{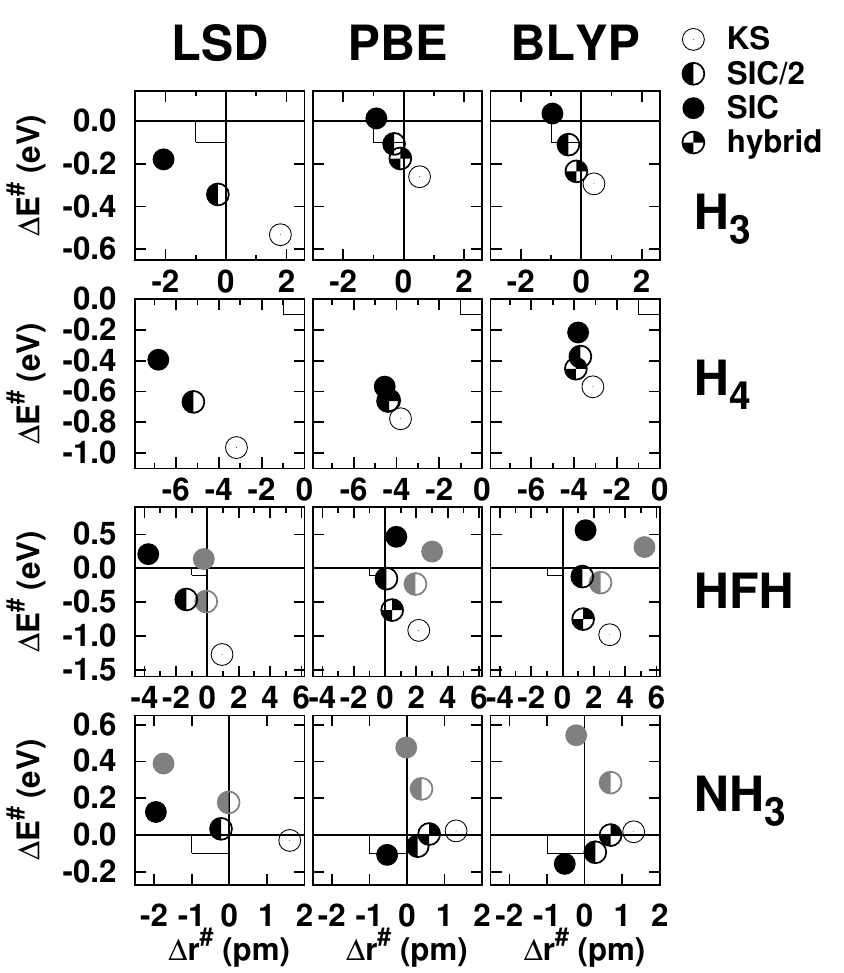}%
\caption{
Energy barrier and bond lengths at the saddle points for the four reactions of Table \ref{tab:barriers}. 
The deviation from reference energy, $\Delta E^\#$ in eV, and bond lengths, $r^\#$ in \AA{}, is shown for the various
functionals. The rectangle, 1 pm by 0.1 eV, at the origin emphasizes the different energy and length scales of the different graphs. For HFH and \ce{NH3} the results based on real orbitals are depicted by grey symbols, for \ce{H4} and \ce{H3} real and complex orbitals give identical results.}
\label{fig:barriers}
\end{figure}

Figure \ref{fig:barriers} shows the deviation of the calculated saddle points from the reference values. For \ce{H4} and \ce{H3}, the barrier height predicted by the uncorrected functionals is always too low, and is increased by SIC/2 and SIC. 
For both reactions, SIC improves the energy barrier but it is still underestimated for \ce{H4}. 
The bond lengths at the saddle point decrease for SIC/2 and further for SIC, as is also found for most molecules. 
Compared to the bond lengths predicted by the
uncorrected functionals, this increases the deviation for \ce{H4} and overcorrects the slightly too large bonds for \ce{H3}. 

The \ce{HFH} barrier is underestimated by the uncorrected functionals and increases with SIC, monotonously from SIC/2 to 
SIC, the latter giving an overcorrection. For the GGA functionals the bond lengths do not change monotonously 
but behave similar to the \ce{F2} bond length discussed above. The bond length decreases for SIC/2 and agrees better with the reference value, but then increases for PBE+SIC, while hardly changing for BLYP+SIC. For the functionals studied here,
the GGA+SIC/2 barriers give the best results, being more accurate than the hybrid functionals. For this reaction, in contrast to the hydrogen barriers, different results are obtained when using real orbitals. Here, the bond length is larger and the barrier height is slightly lower than for complex orbitals.

These three barriers describe bond breaking situations and are underestimated by the 
uncorrected functionals. The ammonia inversion is qualitatively different, as bonds 
merely rearrange and their lengths
at equilibrium and saddle point do not differ significantly. The calculated bond lengths in the planar configuration are elongated by less than 2\,pm compared to the equilibrium structure \cite{Lin02}. The inversion barrier calculated with the uncorrected
functionals is in good agreement with the reference, but the bonds are predicted to be
too long. The bond length decreases and the barrier increases with SIC  for LSD, but for the GGA functionals the barrier is lowered and becomes
underestimated. Using real orbitals, on the other hand, the barrier increases strongly. This qualitatively different effect of SIC can be explained by the 
structure of the `lone pair,' as was done for the methyl radical. For the planar molecule, the orbitals of both spins look very similar to those in Figure 
\ref{fig:CH3_orbitals}. For the complex orbitals, spin up and spin down orbitals are rotated with respect to
each other by 120$^\circ$. In the real case, the spin-up and spin-down orbitals are symmetric with respect to reflection through
the plane. As for \ce{CH3}, this configuration of the real orbitals is higher in energy than that of the complex orbitals, which explains the large difference in the predicted energy barrier.
%

\begin{figure}[tb]
\includegraphics[width=0.85\columnwidth]{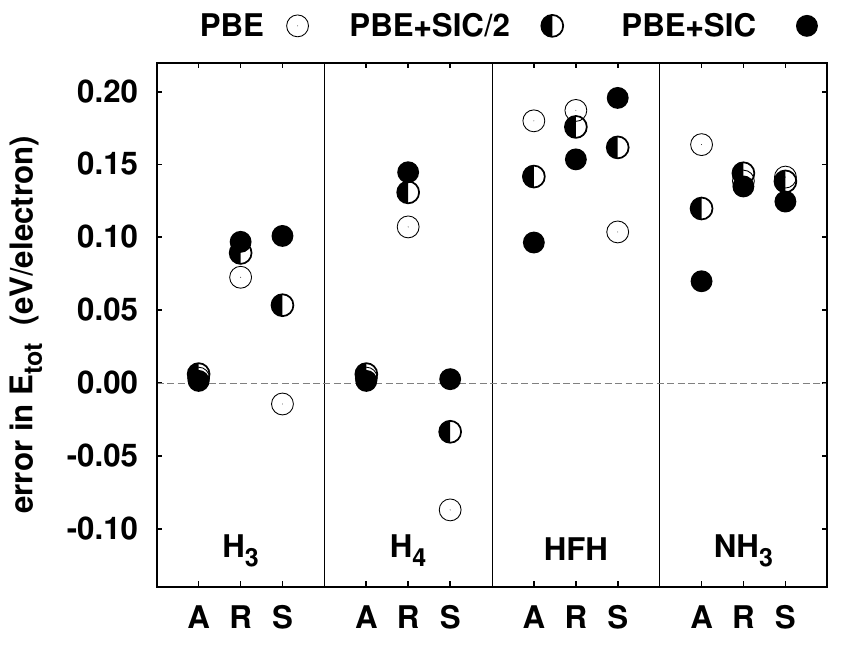}
\caption{Errors in total energy for the PBE-type functionals. For each of the reactions, the errors per electron for the saddle point (S), reactants (R), and separated atoms (A) is shown. The vertical difference between neighboring points corresponds to errors in atomization energy (A-R) and barrier height (R-S).}
\label{fig:total_energy_saddle}
\end{figure}

As for the atomization energy, the errors in the calculated reaction barriers can be analyzed in greater detail by comparison of the errors in total energy of the species involved.
Figure \ref{fig:total_energy_saddle} shows the error per electron of the total energy of the saddle point structure (\textbf{S}), the separated reactants (\textbf{R}) and that of the separated atoms (\textbf{A}), calculated using PBE, PBE+SIC/2, and PBE+SIC. Points 
on the dashed line correspond to a perfect description of the total energy, the vertical difference between neighboring points corresponds to deviations in the atomization energy (A-R) and energy barrier (R-S). 

For the \ce{H3} systems, the PBE energy of the hydrogen atoms is quite accurate
and the saddle point energy is only slightly underestimated. The underestimation of both the barrier height and atomization energy originates mainly from an inaccurate description of the hydrogen molecule. When SIC 
is applied, the energy of the atom is lowered insignificantly and the energy of the molecule is raised slightly, making
both total and atomization energy of the molecule less accurate. 
However, PBE+SIC introduces a large error at the saddle point, which cancels the error in the molecule. For 
SIC/2, the errors introduced are less well balanced and the barrier is still underestimated.

\begin{figure}
\includegraphics[width=0.85\columnwidth]{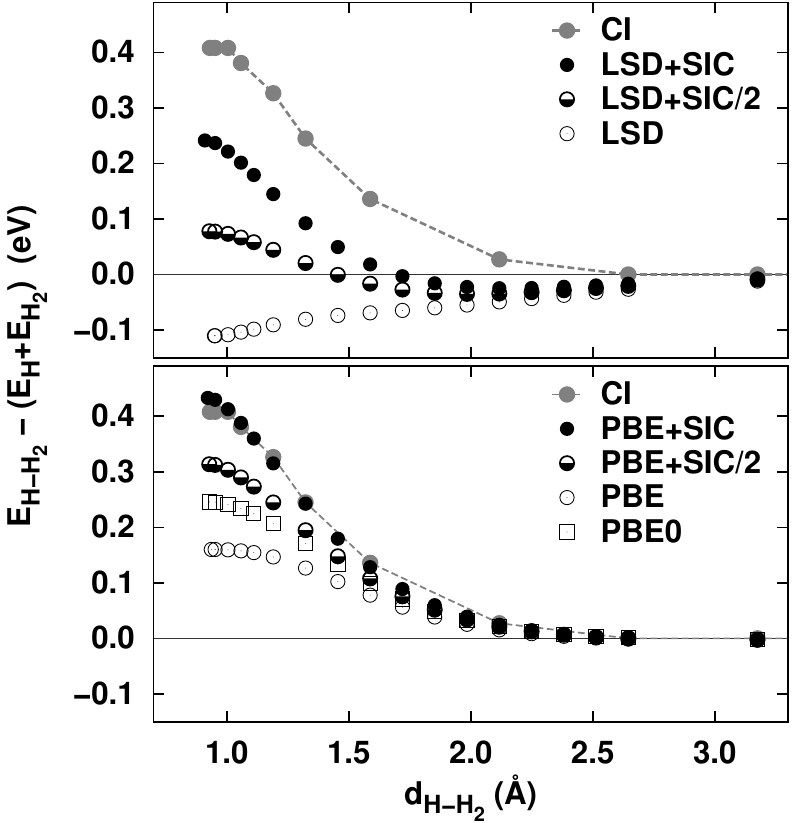}%
\caption{Energy along a path for
the \ce{H + H2 -> H2 + H}  reaction.
The x-axis shows the distance between the two fragments. The \ce{H2} bond length has been relaxed for each separation. LSD+SIC(/2) predicts an energy barrier, but also reveals an intermediate configuration that is
more stable than the separated reactants. Full SIC applied to PBE gives good agreement with the CI results and 
corrects the LSD result enough to avoid the formation of a
 stable hydrogen trimer.}
\label{fig:H3_relaxed}
\end{figure}
The \ce{H4} barrier is severely underestimated by the PBE functional
stemming from equally large errors of opposite sign in the reactants and saddle point structure.  The energy of the saddle point is predicted accurately 
by PBE+SIC. Each optimized orbital is to great extent localized at one of the hydrogen atoms
and neighboring orbitals are of opposite spin. In such a configuration, most of the exchange-correlation energy 
can be expected to be self-interaction energy, in which case PZ-SIC usually performs 
well. In this case the barrier is improved by SIC, but as the energy of the hydrogen molecule
is too high, a still significant error in the barrier remains.

For the atomic reference of HFH and \ce{NH3}, both the error in the total energy and the change when 
applying SIC are dominated by the heavy atoms. The reduction of the error by SIC is smaller for the HF molecule than for the atoms, thus 
making the binding energy less accurate compared to PBE. 
The error in the saddle point energy is increased 
and exceeds that of the reactants.
Here, half SIC gives a better cancellation of error
than both PBE and PBE+SIC. For the \ce{NH3} molecules, the effect is smaller and the barrier height changes only slightly.

For the \ce{H3} reaction, we have 
calculated the energy along a path connecting the saddle point and
the reactants. For a number of distances $d_{\text{H-\ce{H2}}}$, the bond length of the \ce{H2} fragment was relaxed. In Figure \ref{fig:H3_relaxed}, the relaxed energy of the LSD and PBE functionals is shown for several separations of the fragments. The leftmost points correspond to the bond length and 
energy barrier of the \ce{H3} saddle point. 
Increasing values of $d_{\text{H-\ce{H2}}}$ describe the separation into a hydrogen atom and a \ce{H2} molecule. In LSD, the combined system has lower energy than the separated fragments for all separations, the  
energy barrier is negative and \ce{H3} is predicted to be a stable molecule.
LSD+SIC/2 and LSD+SIC increase the \ce{H3} energy and predict a reaction barrier. The height is however still underestimated and at intermediate distances, a system more stable than the reactants is found. In PBE, no intermediate configuration is more stable than the reactants, but the energy difference is globally underestimated. PBE+SIC agrees well with the CI results while the PBE0 hybrid does not describe the system as
well. However, as shown in Figure \ref{fig:total_energy_saddle} the improved description by the PBE+SIC functional results only from a 
better cancellation of errors and not from a better description of the system.

\section{Summary and Conclusion}
The results presented here on the energetics of small molecules provide 
insight into both the strengths and the shortcomings of the Perdew-Zunger self-interaction correction. The qualitatively wrong equilibrium structure of \ce{C2H} predicted by PBE is corrected by SIC. In the case of the \ce{CH3} radical, SIC 
had been found to predict an incorrect geometry, 
but in the present study this
was shown to be an artifact of a restriction to real orbitals.  
This molecule and the \ce{H3} potential energy surface emphasize the importance of using SIC with GGA functionals to obtain more accurate results, as had already been shown in a study of the total energy of atoms~\cite{Klu11}. 
However, not all GGA functionals are suited for the application of SIC. The \ce{F2} molecule is found to be unstable when SIC is applied to the BLYP functional and the errors in total energy are large and unsystematic for the rest of the molecules in the test set as well as the isolated atoms.

Analysis of the total energy revealed that for PBE+SIC, the errors per electron in the atomic systems are reduced and show less fluctuations than that of the molecules, resulting in an unsystematic effect on the atomization energy. The Perdew-Zunger SIC does not in general result in an overcorrection when applied to PBE, as the predicted total energy is still too high. However, an overcorrection in calculated atomization energy is usually observed which can be improved by scaling the SIC. 
The simple scaling scheme of using a constant factor of one-half for all orbitals when applied with PBE
reduces the mean error in atomization energy to less than that of the PBE0 hybrid functional, but still gives 
significant absolute errors. For systems with only a few electrons, this half-SIC approximation does not perform better than full SIC in calculations of
energy barriers.
More flexible scaling \cite{Vyd06,Vyd06a} where full SIC is retained for one-electron systems or isolated orbitals 
could work better for such systems. 
These scaling schemes have so far only been used in combination with real orbitals and their performances would need to be reassessed using complex orbitals.

\vspace{0.5cm}
\section{Acknowledgments}
This work was supported by the Graduate Fellowship Fund of The University of Iceland, the Icelandic Research Fund, the Marie Curie Research Training Network (MCRTN) ``Hydrogen,'' and the 
Nordic Energy Research network ``Solar Fuel.''  The calculations were carried out on the computer clusters ``Sol,'' funded
by the Icelandic Research Fund, and ``Gardar,'' operated by the Nordic High Performance Computing (NHPC) project.



\end{document}